\documentclass[aps,twocolumn,prb,footinbib,showpacs,superscriptaddress,amsmath,amssymb]{revtex4-1}

\usepackage{graphicx}
\usepackage{subfigure}
\usepackage{epsfig} 
\usepackage{dcolumn}
\usepackage{bm}
\usepackage{color}

\begin{document}

\title{Rectifying full-counting statistics in a spin Seebeck engine}

\author{Gaomin Tang}
\affiliation{Department of Physics and the Center of Theoretical and Computational Physics, The University of Hong Kong, Hong Kong, China}
\author{Xiaobin Chen}
\affiliation{Department of Physics and the Center of Theoretical and Computational Physics, The University of Hong Kong, Hong Kong, China}
\author{Jie Ren}
\email{xonics@tongji.edu.cn}
\affiliation{Center for Phononics and Thermal Energy Science, China-EU Joint Center for Nanophononics, Shanghai Key Laboratory of Special Artificial Microstructure Materials and Technology, School of Physics Science and Engineering, Tongji University, 200092 Shanghai, People’s Republic of China}
\author{Jian Wang}
\email{jianwang@hku.hk}
\affiliation{Department of Physics and the Center of Theoretical and Computational Physics, The University of Hong Kong, Hong Kong, China}

\date{\today}

\begin{abstract}
%Traditionally, the rectification and negative differential effect are only studied in the context of heat current behaviors in thermal diodes. 
We investigate the rectification and negative differential effects of full-counting statistics of conjugate thermal spin transport in a spin Seebeck engine. The engine is made by an electron/magnon interface diode driven by a temperature bias, of which the scaled cumulant generating function is formulated in the framework of nonequilibrium Green's function. The strongly fluctuating interfacial electron density of states induced by quantum dot is responsible for these intriguing properties. This work is relevant for designing efficient spin caloritronic engine and can broaden our view in nonequilibrium thermodynamics and the nonlinear phenomenon in quantum transport system.
\end{abstract}

\maketitle

%\section{Introduction}  \label{secI}  
\textit{Introduction--}
Spin caloritronics has emerged as a new field to utilize the excess heat generated in nano-devices to drive spin currents \cite{Bauer2}. Of particular interest is the spin Seebeck effect (SSE) in magnetic insulators, which generates pure spin transfer out of thermal gradient without conducting electron currents, so that opens a new direction to harvest heat in the absence of Joule heating \cite{SSE, SSE1, SSE2, SSE3, SSE4, Jie1, Jie2, Ronetti}. 
 %Of particular interest is the observation of SSE in a magnetic insulator wherein spin waves (a collective motion of spin) can propagate. 
The pure spin is carried by magnons (quantized spin waves) that can transfer energy and spin angular momentum. Besides, magnons have long life-time so that they can propagate over long distances in a ballistic way without being dissipated in some materials due to the low damping~\cite{YIG1, YIG2, YIG3}. %, such as yttrium iron garnet (YIG) \cite{YIG1, YIG2, YIG3}. 

Many proposals have been made in exploiting the application of magnons, such as energy harvester \cite{harvester1}, diodes \cite{Jie2, diode}, transistors \cite{transistor} and logic devices\cite{logic}. 
In particular, the two terminal hybrid system involving both metallic and magnonic reservoir has been proposed \cite{Jie1, Barnas}. In such system, the propagation of spin waves can be converted to spin current in the metallic part and vice versa across the junction under a temperature bias.  
%The electronic type spin current could be converted to a voltage signal through the inverse spin Hall effect so that it can be detected \cite{SSE1,SSE2,SSE3,SSE4}. 
It has been pointed out that strongly fluctuating electron density of states (DOS) in the metallic materials is helpful to obtain the nontrivial rectification and negative differential effects \cite{Jie1, Jie2}, which suggests that quantum dot (QD) structure is a good candidate in achieving these effects. The reduced dimension of QD can also help to increase the figure of merit (FOM) and harvest from the waste heat more efficiently compared to bulk systems when working as a heat engine \cite{QD2, QD3, QD4}.
	
	So far, the rectification and negative differential effect are only studied in the context of mean values of heat and spin currents. For a nanoscale device, fluctuations and even higher-order cumulants of currents can be significant and play an important roles, which could be well described within the framework of full-counting statistics (FCS)\cite{Levitov2, Levitov3, RMP, Hanggi, Ka2, gm2, gm3, gm4}. FCS of a physical quantity is encoded in its cumulant generating function (CGF), which reveals us the higher-order Onsager reciprocity relations \cite{Hanggi, Gaspard1} and the thermodynamic fluctuation symmetry \cite{RMP,Hanggi, Gaspard2} as well. It would illuminate us in better understanding the nonequilibrium thermodynamics and designing efficient heat harvester to study the rectification of higher-order cumulants, fluctuation symmetry, and heat engine performances for the two terminal hybrid system involving both electronic and bosonic reservoirs.
	
	In this work, we investigate the rectification and negative differential effects of FCS in a spin Seebeck engine sandwiched by one electronic and the other magnonic reservoirs. We have formulated a framework to obtain scaled cumulant generating function (SCGF) of both spin and heat currents using nonequilibrium Green's function (NEGF), which does not require week couplings to both reservoirs, and  treats the interfacial electron-magnon interaction to second order. The fluctuation symmetry is obtained and the condition in realizing rectification and negative differential effect is discussed.  The rectification and negative differential effects of cumulants and heat engine performances with respect to temperature reversal are studied in details. 

%to be deleted before submitting	
%	The rest of the article is structured as follows. The model and the Hamiltonian are introduced in Section II. In the following section, SCGF of spin and heat flow is expressed in terms of NEGF. Fluctuation symmetry is investigated. In Section IV, numerical results will be reported. Finally, we summarize our work in section V.

%\section{Model and theoretical formalism}  \label{secII}

\begin{figure}
  \includegraphics[width=2.0in]{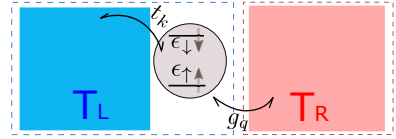} \\
  \caption{(Color online) Schematic illustration of the spin Seebeck engine made by an electron/magnon interface diode. 
%Quantum dot is coupled to the left metallic lead and the right magnonic reservoir. 
Heat flowing across the system can induce spin currents. }
  \label{fig1}
\end{figure}

\textit{Model and theoretical formalism--}
The system is schematically illustrated in Fig.~\ref{fig1} wherein the left electronic compartment interacts with the right magnonic reservoir. The spin current can be generated out of a temperature gradient $\Delta T =T_L-T_R$. The whole system Hamiltonian has contributions from the electrons, magnons and the interaction between them,
\begin{equation} \label{Hamil}
\hat{H} = \hat{H}_{el} + \hat{H}_{mag} + \hat{H}_{em} ,
\end{equation}
The electron Hamiltonian $H_{el}$ is expressed as:
\begin{equation} \label{H_el}
\hat{H}_{el} = \sum_{k\sigma} \epsilon_{k\sigma} a_{k\sigma}^\dag a_{k\sigma} .
\end{equation}
$\hat{H}_{el}$ can incorporate the non-interacting electronic scattering region as well, and we take a single two-level QD in the numerical calculation. The electronic spin-$\sigma$ chemical potential is $\mu_\sigma$ and their difference is defined as the spin bias with the form $\Delta \mu_s = \mu_\downarrow - \mu_\uparrow$ which could be measured by inverse spin Hall effect \cite{SSE2}. The Fermi distribution for the spin-$\sigma$ is
\begin{equation}
f_{L\sigma}(\epsilon_{k\sigma}) = \big\{ \exp[\beta_L(\epsilon_{k\sigma}-\mu_\sigma)]+1 \big\}^{-1},
\end{equation}
with inverse temperature $\beta_L=1/(k_B T_L)$.

	The right insulating magnetic compartment can be described by a Heisenberg lattice 
\begin{equation*}
\hat{H}_{mag} = -J\sum_{\langle i,j\rangle} \left[ \frac{1}{2}S_{i+} S_{j-} + \frac{1}{2}S_{i-} S_{j+} + S_{iz} S_{jz} \right], 
\end{equation*}
where $S_{j\pm}$ is the localized raising (lowering) spin operator at site $j$, $S_{jz}$ is the spin operator in the $z$ direction, and $J$ denotes the exchange coupling strength. Using Holstein-Primakoff transformation \cite{HP, Mahan},  
$S_{j+} = \sqrt{2S_0 - b_j^\dag b_j}b_j, \ S_{j-} = b_j^\dag\sqrt{2S_0 - b_j^\dag b_j}, \ S_{jz} = S_0 - b_j^\dag b_j$ with localized spin length $S_0$, the spin operators are mapped into bosonic magnons. At large spin limit \cite{Jie1,harvester1,QD4}, i.e., $2S_0 \gg \langle b_j^\dag b_j\rangle$, we have the approximation $S_j^- \approx \sqrt{2S_0} b_j^\dag$ and $S_j^+ \approx \sqrt{2S_0}b_j$. The right magnetic insulator can be approximated by the free magnon gas \cite{Jie1}: 
\begin{equation}
\hat{H}_{mag} \approx \sum_q \hbar\omega_q b_q^\dag b_q + {\rm const} ,
\end{equation}
after a Fourier transform into the momentum space, where the dispersion of $\omega_q$ depends on the material details. The magnonic reservoir obeys the Bose-Einstein distribution $N_R(\omega)=1/(e^{\beta_R\omega}-1)$, where the chemical potential of magnons is set to zero due to the ferromagnetic insulator phase considered here \cite{Magnetic_RC_Circuit}.
The interfacial electron-magnon interaction Hamiltonian bears the form \cite{Jie1, Mahan},
\begin{equation} 
\hat{H}_{em} = - \sum_{kk'q} g_q \left[\gamma_{k\uparrow}^* \gamma_{k'\downarrow} a_{k\uparrow}^\dag a_{k'\downarrow} b_q^\dag + {\rm H.c.} \right] \delta_{k-k'=q} ,
\end{equation}
with the delta function $\delta_{k-k'=q}$ ensuring energy and spin angular momentum conservation.
The interaction describes magnon assisted spin flip by absorbing or emitting magnons with frequency $\omega_q$. The effect from the electronic DOS is incorporated in $\gamma_{k\sigma}$. The electron-magnon interaction strength $g_q$ is assumed to be weak and will be treated perturbatively in getting SCGF. For simplicity, one can assume $g_q$ is energy independent with $g_q= g$.

%%%%% FCS for spin and heat currents

	Spin (s) and heat (h) current operators are defined as the change rate of the electronic spin in the left compartment and heat in the right reservoir, respectively, with the forms $\hat{I}_s(t)= -d_t \hat{N}_s$, $\hat{I}_h(t)= -d_t \hat{H}_R$, and $d_t$ being the total differential with respect to time in Heisenberg picture. 
Here, $\hat{N}_s=\frac{1}{2}\sum_k (a_{k\downarrow}^\dag a_{k\downarrow} - a_{k\uparrow}^\dag a_{k\uparrow})$ is the operator of number of spin in the left compartment and $\hat{H}_R=\sum_q \hbar\omega_q b_q^\dag b_q$.
Using random phase approximation (RPA), that is treating the interaction strength $g$ between the electrons and magnon modes perturbatively to the second order \cite{Ka2}, we could obtain SCGF for spin and heat current (with counting fields $\lambda_s$ and $\lambda_h$) expressed as (See Supplemental Material \cite{supp} for details of derivation),
\begin{align} \label{SCGF}
{\cal G}(\lambda_s, & \lambda_h) = -\int \frac{d\omega}{2\pi} \ln \bigg\{ 1-\int \frac{dE}{2\pi} A(E_-,E_+) \notag \\
\Big[ &(e^{+i\lambda_s+i\hbar\omega\lambda_h}-1)  f_{\downarrow\rightarrow\uparrow}(E_+, E_-) (N_R(\omega)+1)  \notag \\
 +&(e^{-i\lambda_s-i\hbar\omega\lambda_h}-1) f_{\uparrow\rightarrow\downarrow}(E_-, E_+) N_R(\omega) \Big] \bigg\},
\end{align}
with
\begin{equation}
A(E_-,E_+)=\frac{J_R(\omega) J_{L\uparrow}(E_-)J_{L\downarrow}(E_+)}{\big|[\Sigma_R^r(\omega) + F^r(\omega)\big|^2} .
\end{equation} 
Here we have defined $f_{\uparrow\rightarrow\downarrow}(E_-,E_+) = f_{L\uparrow}(E_-) [1- f_{L\downarrow}(E_+)]$ and $f_{\downarrow\rightarrow\uparrow}(E_+,E_-) = f_{L\downarrow}(E_+) [1- f_{L\uparrow}(E_-)]$ with $E_{\pm}=E \pm \omega/2$.
$J_{L\sigma}(E) = 2\pi g\sum_k |\gamma_{k\sigma}|^2 \delta(E-\epsilon_{k\sigma})$ is the
the spin-$\sigma$ electronic spectral density. The retarded magnonic self-energy due to the dissipation is expressed as $\Sigma_R^r = -iJ_R(\omega)/2$. Without loss of generality, the magnonic reservoir spectral function is considered to be Ohmic and expressed as $J_R(\omega) = \pi\alpha\omega e^{-\omega/\omega_c}$, where $\alpha$ is the dimensionless dissipation strength and $\omega_c$ is the cut-off frequency \cite{Weiss, supp}. The expression of retarded electron-hole propagator is,
\begin{equation}
F^r(\omega)= i \int\frac{d\omega_1}{2\pi} \frac{F^>(\omega_1)-F^<(\omega_1)}{\omega -\omega_1 +i0^+}.
\end{equation}
with 
\begin{align*} 
F^<(\omega) &= -i\int \frac{dE}{2\pi}\Big[ J_{\uparrow}(E_-) J_{\downarrow}(E_+) f_{\downarrow\rightarrow\uparrow}(E_+, E_-) \Big] ,  \\
F^>(\omega) &= -i\int \frac{dE}{2\pi}\Big[ J_{\uparrow}(E_-) J_{\downarrow}(E_+) f_{\uparrow\rightarrow\downarrow}(E_-, E_+) \Big]  .
\end{align*}
The physical picture of SCGF is clear. The first contribution of Eq.~\eqref{SCGF} describes that the spin-down electrons with energy $E+\hbar\omega/2$ flip to the spin-up states with energy $E-\hbar\omega/2$ by emitting a magnon with energy $\hbar\omega$ into the magnonic reservoir. The second one describes a reverse process that the spin-up electrons with energy $E-\hbar\omega/2$ flip to the spin-down states with energy $E+\hbar\omega/2$ via absorbing a magnon carrying energy $\hbar\omega$.

	Applying the identity $f_{\uparrow\rightarrow\downarrow}(E_-,E_+) = e^{\beta_L(\omega-\Delta\mu_s)} f_{\downarrow\rightarrow\uparrow}(E_+,E_-)$ in Eq.~\eqref{SCGF}, we can obtain the following fluctuation symmetry relation \cite{RMP, Hanggi},
\begin{equation} \label{FT}
{\cal G}(\lambda_s, \lambda_h)
= {\cal G}(-\lambda_s -i\beta_L\Delta\mu_s, -\lambda_h+i(\beta_L-\beta_R)).
\end{equation}
If $J_{L\sigma}(E)$ were flat near Fermi energy, $A(E_-,E_+)$ inside the integral in Eq.~\eqref{SCGF} would be treated as a constant. By applying the equality $\int \frac{dE}{2\pi} f_{\downarrow\rightarrow\uparrow}(E_+, E_-)=(\hbar\omega-\Delta\mu_s)N_L(\hbar\omega-\Delta\mu_s)$ and assuming $\Delta\mu_s \rightarrow 0$, the SCGF would be almost symmetric by reversing the temperature gradient $\Delta T$ with a constant electron DOS and so are the cumulants of currents, that is, ${\cal G}(\lambda_s, \lambda_h)={\cal G}(-\lambda_s, -\lambda_h ; L\leftrightarrow R)$. This is usually satisfied in the pure fermionic system or bosonic system, but is not satisfied in the electron-magnon hybrid system discussed here.  {\it Therefore, strongly fluctuating electron DOS is essential to have asymmetric behavior with respect to reversing the temperature gradient and rectify FCS and heat engine performances of the spin Seebeck engine.} 
	The spin current in the electronic reservoir and heat current in the magnonic reservoir could be derived by taking the first order derivative of ${\cal G}(\{\lambda\})$ with respect to $\lambda_s$ and $\lambda_h$, respectively at $\lambda_s=\lambda_h=0$, that is, $I_{s/q} =\partial {\cal G}(\{\lambda\}) /\partial (i\lambda_{s/q})|_{\lambda_s=\lambda_h=0}$. Then,
\begin{align} \label{currents}
&I_{s(h)} =  \int\frac{d\omega}{2\pi}\int\frac{dE}{2\pi}\Big\{ 
(\hbar\omega)^\alpha A(E_-,E_+) \times  \notag \\
& \left[ f_{\downarrow\rightarrow\uparrow}(E_+,E-) (N_R(\omega)+1) 
- f_{\uparrow\rightarrow\downarrow}(E_-,E_+) N_R(\omega)  \right] \Big\}  ,
\end{align}
where $\alpha=0$ for spin current $I_s$, and $\alpha=1$ for heat current $I_h$.

	We now discuss the device working as a thermoelectric engine with the output spin power $P=|I_s\Delta\mu_s|$. When $\Delta T = T_L-T_R <0$, heat current flows from the right reservoir to drive the spin current, the heat current calculated from Eq.~\eqref{currents} is negative and the efficiency should be defined as $\eta =P/|I_h|$. One should note that the spin bias which is defined as $\Delta\mu_s=\mu_\downarrow-\mu_\uparrow$ should be chosen as positive $\Delta\mu_s >0$ under $\Delta T <0$, otherwise the device works as a unphysical dud engine where output spin power heats both reservoirs.
In the case of $\Delta T>0$, i.e., $T_L > T_R$, the left reservoir is the heat source with spin bias $\Delta\mu_s <0$ and heat current is $I_{Lh}=|I_h|+P$ using energy conservation law, so that the efficiency has the definition of the form $\eta =P/(|I_h|+P)$.

\begin{figure}
  \includegraphics[width=3.7in]{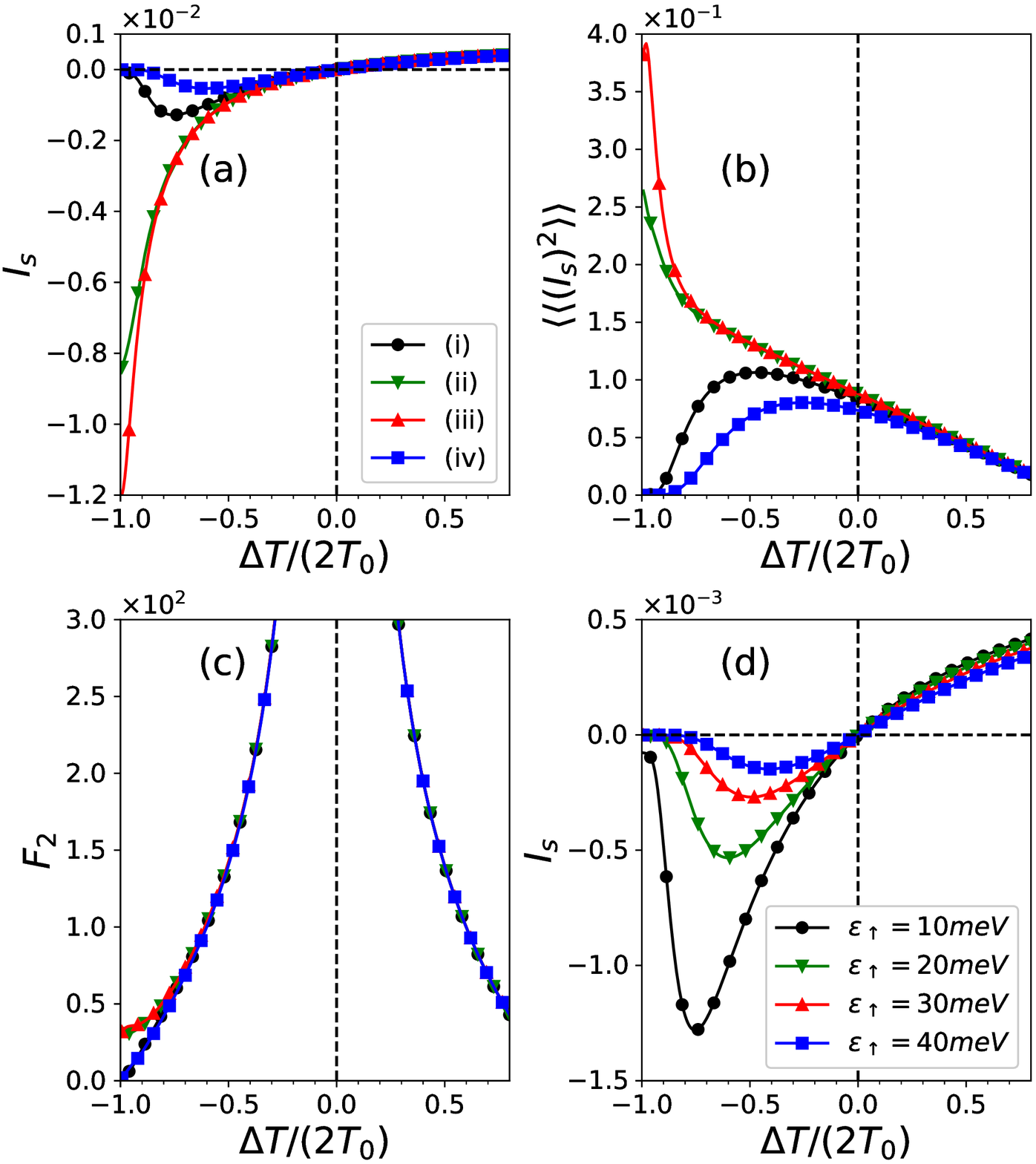} \\
  \caption{(Color online) (a) Spin current $I_s$, (b) spin current noise $\langle\langle (I_s)^2\rangle\rangle$, and (c) Fano factor $F_2$ as a function of $\Delta T/(2T_0)$ by varying $\epsilon_\sigma$: (i) $\epsilon_\downarrow =15meV$, $\epsilon_\uparrow =10meV$; (ii) $\epsilon_\downarrow =5meV$, $\epsilon_\uparrow =0$; (iii) $\epsilon_\downarrow =3meV$, $\epsilon_\uparrow =-2meV$; (iv) $\epsilon_\downarrow =-20meV$, $\epsilon_\uparrow =-25meV$. (d) Spin current $I_s$ as a function of $\Delta T/(2T_0)$ by varying $\epsilon_\uparrow$ with $\epsilon_\downarrow=\epsilon_\uparrow+5meV$.  $\Gamma =4meV$.  }
  \label{fig2}
\end{figure}

%\section{Numerical results}   \label{secIII}
\textit{Numerical results--}
In order to get a fluctuating electron DOS, one can, for example, insert a QD between the electronic and magnonic reservoir. Hamiltonian of the electronic part is then expressed as
\begin{equation*}
\hat{H}_{el} =\sum_\sigma \epsilon_{\sigma} d_\sigma^\dag d_\sigma + \sum_{k\sigma}\Big[\epsilon_{k\sigma} c_{k\sigma}^\dag c_{k\sigma} + (t_{k\sigma}c_{k\sigma}^\dag d_\sigma + {\rm H.c.})\Big] , 
\end{equation*}
with the first term describing QD, the second term electronic reservoir and the rest for their coupling. $\epsilon_{\sigma}$ are the two energy levels inside QD. We consider a large quantum dot so that the Coulomb interaction effect can be neglected. $\epsilon_{k\sigma}$ and $t_{k\sigma}$ are the energy level of state $k\sigma$ in the electronic reservoir and its coupling strength with the spin-$\sigma$ in QD.
	The electron Hamiltonian $\hat{H}_{el}$ with QD can be diagonalized and rewritten in the form of  Eq.~\eqref{H_el}
through the linear transformation \cite{Bijay, Segal}
\begin{equation}
d_\sigma = \sum_k \gamma_{k\sigma} a_{k\sigma} , \ \ \ c_{k\sigma} = \sum_{k'} \eta_{kk'\sigma} a_{k'\sigma} 
\end{equation}
where the dimensionless coefficients are
\begin{equation*}
\gamma_{k\sigma} = \frac{t_{k\sigma}}{\epsilon_{k\sigma}-\epsilon_{\sigma} -\Sigma_{k\sigma} } ,  \ \
\eta_{kk'\sigma} = \delta_{kk'} - \frac{t_{k\sigma}\gamma_{k'\sigma}}{\epsilon_{k\sigma}-\epsilon_{k'\sigma}+i0^+} ,
\end{equation*}
with the self-energy function $\Sigma_{k\sigma}=\sum_{k'} t_{k'\sigma}^2/(\epsilon_k-\epsilon_{k'}+i0^+)$. The electronic spectral densities in the left compartment (electronic reservoir plus QD) is Lorentzian shaped with the form \cite{Bijay}
\begin{equation} \label{J_L}
J_{L\sigma}(E)= 2\pi g\sum_k |\gamma_{k\sigma}|^2 \delta(E-\epsilon_{k\sigma})= \frac{g\Gamma_\sigma}{(E-\epsilon_\sigma)^2+\Gamma_\sigma^2/4} ,
\end{equation}
where the electronic coupling strength $\Gamma_\sigma=2\pi\sum_k |t_{k\sigma}|^2 \delta(E-\epsilon_{k\sigma})$ is assumed energy independent.

	During the numerical calculation, the coupling strengths for the spin up and down electrons between the electronic reservoir and QD are assumed to be energy independent and equal, i.e., $\Gamma_{\uparrow}=\Gamma_{\downarrow}=\Gamma$. The energy level of spin up is set to be larger than the that of spin down in the QD, $\epsilon_\downarrow > \epsilon_\uparrow$. The dimensionless dissipation strength and the cut-off frequency of the magnonic reservoir is chosen as $\alpha=0.2$, and $\omega_c=20meV$, respectively.
The temperatures of the reservoirs are set as $T_{L,R}=T_0\pm \Delta T/2$ with $T_0=300K$, so that when the normalized temperature gradient $\Delta T/(2T_0)<0$, the left lead is cooler. We denote the average of spin up and down chemical potential as $\mu_0=(\mu_\uparrow + \mu_\downarrow)/2$. The spin currents are displayed in atomic unit.

	In Fig.~\ref{fig2}, we plot the spin current $I_s$ [panel (a)], spin current noise $\langle\langle (I_s)^2\rangle\rangle$ [panel (b)], and Fano factor $F_2=\langle\langle (I_s)^2\rangle\rangle/I_s$ [panel (c)] as a function of the normalized temperature gradient $\Delta T/(2T_0)$ by varying $\epsilon_\sigma$. The spin up and down chemical potentials are both set to zero, i.e., $\mu_\uparrow = \mu_\downarrow =0$. The rectification effect where the spin current are asymmetric with respect to temperature gradient reversal can be clearly identified in Fig.~\ref{fig2}. In the cases where the two levels of QD are either both above or below the chemical potential ($\epsilon_\downarrow > \epsilon_\uparrow > 0$ in case (i) or $0>\epsilon_\downarrow > \epsilon_\uparrow$ in case (iv) in Fig.~\ref{fig2}), we could observe the negative differential SSE that the spin current decrease or even vanish with the increasing temperature gradient $\Delta T$. This could be understood as follows: when the right reservoir is hotter, i.e., $\Delta T <0$, the process that spin up electron absorbs energy to flip to the spin down state dominates. Near the linear response regime, increasing the magnitude of temperature gradient which corresponds to decreasing $T_L$ leads to increased heat current to assist the spin flip process. When the two levels are both above (below) the chemical potential, the electrons on these levels can be depleted (occupied) by decreasing $T_L$ further. Compared to the increasing $N_R(\omega)$, $f_{\uparrow\rightarrow\downarrow}(E_-,E_+) = f_{L\uparrow}(E_-) [1- f_{L\downarrow}(E_+)]$ is severely suppressed and so is the flipping process from spin up to down state \cite{Jie1}. {\it Thus, the essential ingredient to get negative differential SSE is that the two levels are on one side of the chemical potential.}
We can also observe from Fig.~\ref{fig2}(b) that spin current noise is asymmetric with respect to temperature gradient which signifies rectification as well. 
If $J_{L\sigma}(E)$ is flat near $\epsilon_\sigma$, spin current noise is an even function of $\Delta T$, which is not satisfied in Fig.~\ref{fig2}(b). For case (ii) and (iii), noise are monotonic, while for case (i) and (iv), decrease of noise with increasing the magnitude of $\Delta T$ ($\Delta T/(2T_0)<-0.5$) is due to the severely suppressed number of electrons involving in spin flip. One can observe that the Fano factors shown in Fig.~\ref{fig2}(c) are relatively symmetric and almost the same for all cases, except in the region $\Delta T/(2T_0)\rightarrow -1$. 

	In Fig.~\ref{fig2}(d), spin current $I_s$ versus $\Delta T$ by varying $\epsilon_\uparrow$, which is above the chemical potential, is plotted. With increasing $\epsilon_\uparrow$ away from the electronic chemical potential, negative differential points shift towards $\Delta T=0$, while negative differential and asymmetric behaviors become less obvious. This is due to the fact that the number of electrons near $\epsilon_\sigma$ for spin flip is becoming less with increasing $\epsilon_\uparrow$. We conclude that the differential points can be adjusted by tuning QD levels.

\begin{figure}
  \includegraphics[width=3.7in]{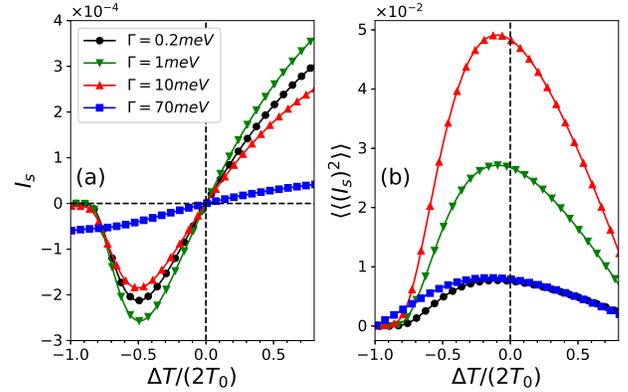} \\
  \caption{(Color online) (a) Spin current, and (b) spin current noise as a function of $\Delta T/(2T_0)$, by varying electronic coupling strength $\Gamma$, with $\epsilon_\uparrow=30meV$ and $\epsilon_\downarrow=35meV$.  }
  \label{fig3}
\end{figure}

	The spin currents and spin current noises versus $\Delta T/(2T_0)$ by varying the electronic coupling strength $\Gamma$ with $\epsilon_\uparrow=30meV$ and $\epsilon_\downarrow=35meV$ are shown in Fig.~\ref{fig3}(a) and (b), respectively.
With increasing $\Gamma$, the spin current increases first and then decreases. Since when $\Gamma$ is very small, $\Gamma \approx 0.2meV$, the electronic spectral density is quite small so that the spin flip is limited by $\Gamma$ and the spin current will increase with increasing $\Gamma$. However when $\Gamma$ is large, increasing $\Gamma$ reduces the spin current because of the level broadening. The negative differential SSE disappears once $\Gamma$ is too large, since electron DOS becomes flatter near $\epsilon_\sigma$ with increasing $\Gamma$. 
Spin current noise is getting symmetric with increasing $\Gamma$ due to electron DOS flattening, this holds as well for the case of $\epsilon_\downarrow >0 > \epsilon_\uparrow$, which is not shown here.

\begin{figure}
  \includegraphics[width=3.7in]{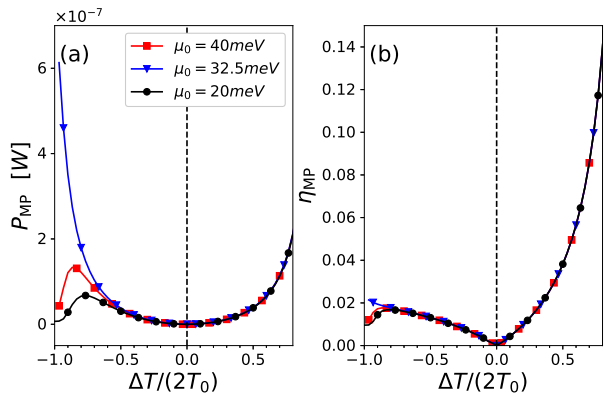} \\
  \caption{(Color online) (a) Maximum output spin power $P_{\rm MP}$, and (b) efficiency $\eta_{\rm MP}$ at maximum power as a function of $\Delta T/(2T_0)$ by varying $\mu_0$ with $\epsilon_\uparrow=30meV$, $\epsilon_\downarrow=35meV$ and $\Gamma=4meV$.  }
  \label{fig4}
\end{figure}

	For a heat engine, maximum output power and maximum efficiency cannot be satisfied at the same time. People usually care about the efficiency $\eta_{\rm MP}$ when the system has a maximum power $P_{\rm MP}$ or the output power under a maximum efficiency at a given $\Delta T$ \cite{MP2,MP3,PED}. 
In Fig.~\ref{fig4}, we plot the maximum output spin power $P_{\rm MP}$, and efficiency $\eta_{\rm MP}$ at maximum power as a function of $\Delta T/(2T_0)$ by varying the average chemical potential $\mu_0$. We observe that both $P_{\rm MP}$ and $\eta_{\rm MP}$ are asymmetric with respect to the temperature reversal. When $\mu_0$ is above or below both levels $\epsilon_\sigma$ of the QD, we can observe the negative differential effect which is also due to the severely decreased electron number  near $\epsilon_\sigma$ with decreasing $T_L$.

%\section{Conclusion}   \label{secIV}
\textit{Conclusion--}
In this work we have uncovered the rectification and negative differential effects of FCS and heat engine performances in a electron/magnon diode under a temperature gradient. The SCGF of spin and heat flow of the system were obtained in the framework of NEGF. Spin and heat currents expressions were derived and fluctuation symmetry was verified from the SCGF. We verified that the strongly fluctuating electron DOS induced by QD is crucial to get these intriguing effects. In principle, multiple QDs can be sandwiched between metallic and magnonic reservoirs to form multiple transport channels in parallel to enhance transport. Without loss of generality, we focus on the single two-level QD here, as a demonstration of the intriguing properties studied.
The hybrid two terminal setup involving both electronic and bosonic reservoirs exhibits a perfect testing ground for some nontrivial phenomena in nonequilibrium thermodynamics.
%Rectification and negative differential effect were also observed in maximum output spin power and efficiency at maximum power which are used to characterize performances of a thermoelectric engine. 
\\
\\
%\begin{acknowledgments}
G. Tang, X. Chen and J. Wang are financially supported by the General Research Fund (Grant No. 17311116), the University Grant Council (Contract No. AoE/P-04/08) of the Government of HKSAR, and NSF-China (Grant No. 11374246). J. Ren acknowledges support from the National Youth 1000 Talents Program in China, and the 985 startup Grant (205020516074) at Tongji University.
%\end{acknowledgments}

\end{document}

%\preprint{APS/123-QED}

\title{Supplemental Material for "Rectifying full-counting statistics in a spin Seebeck engine"}

\author{Gaomin Tang}
\affiliation{Department of Physics and the Center of Theoretical and Computational Physics, The University of Hong Kong, Hong Kong, China}
\author{Xiaobin Chen}
\affiliation{Department of Physics and the Center of Theoretical and Computational Physics, The University of Hong Kong, Hong Kong, China}
\author{Jie Ren}
\email{xonics@tongji.edu.cn}
\affiliation{Center for Phononics and Thermal Energy Science, School of Physics Science and Engineering, Tongji University, 200092 Shanghai, People’s Republic of China}
\author{Jian Wang}
\email{jianwang@hku.hk}
\affiliation{Department of Physics and the Center of Theoretical and Computational Physics, The University of Hong Kong, Hong Kong, China}

\date{\today}

\begin{abstract}
In this supplementary material, we provide the detailed derivation of the scaled cumulant generating function (SCGF) of the spin Seebeck engine in the framework of nonequilibrium Green's function. The expression for spin and heat current and spin current noise are derived from the SCGF. 
\end{abstract}

\maketitle
%\section*{Derivation of generating function}

The partition function of the whole system reads as
\begin{equation}
Z(t) = \int d[\bar{\phi}\phi\bar{\Psi}\Psi] e^{i{\cal S}} .
\end{equation}
The action of the system Hamiltonian consists of three parts
\begin{equation}
{\cal S} = {\cal S}_{el} + {\cal S}_{mag} + {\cal S}_{em} , 
\end{equation}
and has the following expression \cite{Ka2}, 
\begin{align}
&{\cal S}=\int_C dt \bigg\{ \sum_{k\sigma}\bar{\phi}_{k\sigma} g_{k\sigma}^{-1} \phi_{k\sigma} + \sum_q \bar{\Psi}_q D_q^{-1} \Psi_q  \notag \\
&- \sum_{kk'q} g_q\left[ \gamma_{k\uparrow}^*\gamma_{k'\downarrow}\bar{\phi}_{k\uparrow}\phi_{k'\downarrow}\bar{\Psi}_q + \gamma_{k'\downarrow}^*\gamma_{k\uparrow}\bar{\phi}_{k'\downarrow}\phi_{k\uparrow}\Psi_q \right] \delta_{k-k'=q} \bigg\} .
\end{align}
In the above expression, $\bar{\phi}_{k\sigma}$ and $\phi_{k\sigma}$ are the fermionic Grassmann variables for the left electronic compartment, $\Psi_q$ and $\bar{\Psi}_q$ are the complex numbers corresponding to magnonic creation or annihilation operators. $g_{k\sigma}^{-1}$ and $D_q^{-1}$ are the inverse electronic and magnonic Green's functions. Delta function $\delta_{k-k'=q}$ ensures energy and spin angular momentum conservation. The integration is over the Keldysh contour $C$. The action could be further simplified by integrating out the fermionic Grassmann variables so that the partition function has the form as
\begin{align}
&Z(t) =\int d[\bar{\Psi}_q\Psi_q] \exp \bigg[\int_C dt \sum_q \bar{\Psi}_q D_q^{-1} \Psi_q \notag \\
&+ {\rm Tr}\ln\bigg(I- g^2 \sum_{kk'q} \bar{\Psi}_q |\gamma_{k\uparrow}|^2 |\gamma_{k'\downarrow}|^2 g_{k\uparrow} g_{k'\downarrow} \Psi_q \delta \bigg) \bigg] ,
\end{align}
where we have assumed the electron-magnon interaction strength $g_q$ is energy independent and used shorthand notation $\delta = \delta_{k-k'=q}$. 
We can as well write the dependencies of the Keldysh branches explicitly out,
\begin{align}
&Z(t) = \int d[\bar{\Psi}_q \Psi_q] \exp \Bigg[\int dt \sum_q \bar{\Psi}_q^a (D_q^{-1})^{ab} \Psi_q^b  \notag \\
&+ {\rm Tr}\ln\bigg(I- g^2 \sum_{kk'q} \bar{\Psi}_q^a\gamma^a |\gamma_{k\uparrow}|^2 g_{k\uparrow} \gamma^b |\gamma_{k'\downarrow}|^2 g_{k'\downarrow} \Psi_q^b \delta \bigg) \Bigg] ,
\end{align}
with the Keldysh branch index $a,b=+,-$ and 
\begin{equation}
\gamma^+ = \begin{bmatrix}
1 & 0 \\  0 & 0
\end{bmatrix} , \qquad
\gamma^- = \begin{bmatrix}
0 & 0 \\  0 & -1
\end{bmatrix}  .
\end{equation}
In order to consider the dissipation of magnon mode $\omega_q$ due to magnonic reservoir, one can introduce self-energy $\Sigma_{qR}$ for each magnon mode so that the partition function can be written as
\begin{align}
&Z(t) = \int d[\bar{\Psi}_q \Psi_q] \exp \Bigg[\int dt \sum_q \bar{\Psi}_q^a (\underline{D}_q^{-1}-\Sigma_{qR})^{ab} \Psi_q^b  \notag \\
&+ {\rm Tr}\ln\bigg(I- g^2 \sum_{kk'q} \bar{\Psi}_q^a\gamma^a |\gamma_{k\uparrow}|^2 g_{k\uparrow} \gamma^b |\gamma_{k'\downarrow}|^2 g_{k'\downarrow} \Psi_q^b \delta \bigg) \Bigg] .
\end{align}
$\underline{D}_q$ is the magnon Greens' function without taking the reservoir effect into account. %Since different magnon modes $\omega_q$ do not interact with each other, $Z(t)$ is a product with respect to different modes under the restriction $q=k-k'$.

	Treating the interaction strength $g$ between electrons and magnon mode $\omega_q$ perturbatively to second order, that is expanding ${\rm Tr}\ln(I-\cdots)$ to the first order, we can then perform random phase approximation and get the partition in the form of a Fredholm determinant in the time domain, \cite{Ka2}
\begin{align}
 Z(t) = \frac{1}{\det \left[-i(\underline{D}_q^{-1} -\Sigma_{qR} - F_q)\right]}.
\end{align}
where the determinant is over both the time domain and wave vector $q$.
The Keldysh components of electron-hole propagator have the expressions
\begin{equation}
F_q^{ab}(t_1, t_2) = -i g^2 \sum_{kk'q} |\gamma_{k'\downarrow}|^2  |\gamma_{k\uparrow}|^2 
g_{k\uparrow}^{ab} (t_1-t_2) g_{k'\downarrow}^{ba} (t_2-t_1)\delta ,
\end{equation}
with $a,b=+,-$.
The lesser and greater propagator is related to $F^{+-}$ and $F^{-+}$ with the relations $F^<=-F^{+-}$   and $F^>=-F^{-+}$. $(\underline{D}_q^{+-})^{-1}$ and $(\underline{D}_q^{-+})^{-1}$ can be treated zero in the time domain \cite{Ka2}.

The partition obtained above are Fredholm determinant in the time domain without taking counting fields into consideration. In order to study full-counting statistics (FCS) of the spin Seebeck engine, we introduce counting fields $\lambda_s$ and $\lambda_h$ for spin and heat current, respectively. 
Taking the counting fields $\lambda_s$ and $\lambda_h$ into consideration, we can have the expression of the normalized generating function in the time domain as \cite{gm2,RMP}
\begin{widetext}
\begin{equation}
{\cal Z}(t, \{\lambda \}) = \frac{\det\begin{bmatrix}
(\underline{D}_q^r)^{-1}-\Sigma_{qR}^t -F_q^t  & -\Sigma_{qR}^{+-} - F_q^{+-}  \\
-\Sigma_{qR}^{-+} - F_q^{-+} &  -(\underline{D}_q^a)^{-1}-\Sigma_{qR}^{\bar{t}} - F_q^{\bar{t}}
\end{bmatrix}}{\det\begin{bmatrix}
(\underline{D}_q^r)^{-1}-\Sigma_{qR}^t -F_q^t  &  -\widetilde{\Sigma}_{qR}^{+-} - \widetilde{F}_q^{+-}  \\
-\widetilde{\Sigma}_{qR}^{-+} -\widetilde{F}_q^{-+} & -(\underline{D}_q^a)^{-1}-\Sigma_{qR}^{\bar{t}} - F_q^{\bar{t}}
\end{bmatrix}}  ,
\end{equation}
\end{widetext}
with
\begin{align} \label{F_time}
\widetilde{F}_q^{ab}(t_1,t_2) &=  F_q^{ab}(t_1, t_2) e^{i(a-b)\lambda_s/2} ,  \notag \\
\widetilde{\Sigma}_{qR}^{ab} (t_1,t_2) &= \Sigma_{qR}^{ab} (t_1-a\lambda_h/2,t_2-b\lambda_h/2) .
\end{align} 
We can observe from the generating function that electron-hole propagator $F$ is perturbatively coupled to the magnon Green's function $D_q^{-1}$ up to second order of electron-magnon coupling strength $g$. 
In the long time limit, we can have the scaled cumulant generating function (SCGF) 
\begin{equation} \label{ZZ1}
{\cal G}(\lambda_s,\lambda_h) = \lim_{t\rightarrow\infty} \frac{{\cal Z}(t, \{\lambda \})}{t} 
= -\int \frac{d\omega}{2\pi} \ln \det \left[ \frac{D^{-1}_\lambda(\omega)}{D^{-1}_{\lambda=0}(\omega)} \right] ,
\end{equation}
where the determinant is over Keldysh space and wave vector $q$.
The inverse magnon Green's function matrix with counting fields $D^{-1}_{\lambda}(\omega)$ could be explicitly written in the Keldysh space as
\begin{equation} \label{Dyson}
D^{-1}_{\lambda}(\omega) = \begin{bmatrix}
(\underline{D}_q^r)^{-1}-\Sigma_{qR}^{t}-F_q^{t} & \widetilde{\Sigma}_{qR}^{<}+\widetilde{F}_q^{<} \\
\widetilde{\Sigma}_{qR}^{>}+\widetilde{F}_q^{>} & -(\underline{D}_q^a)^{-1}-\Sigma_{qR}^{\bar{t}}-F_q^{\bar{t}} 
\end{bmatrix} .
\end{equation}
The retarded and advanced free-magnon Green's function $D_q^{r,a}(\omega)$ are given by
\begin{equation}
\underline{D}_q^r(\omega) = [\underline{D}_q^a(\omega)]^* =\frac{1}{\omega-\omega_q+i0^+} .
\end{equation}
$\widetilde{\Sigma}_{qR}$ is the self-energy due to the right magnonic reservoir involving heat current counting field $\lambda_h$. Its lesser and greater component, respectively, has the form
\begin{align}
\widetilde{\Sigma}_{qR}^<(\omega) &= -i J_{qR}(\omega) N_R(\omega) e^{-i\hbar\omega\lambda_h} ,  \notag \\
\widetilde{\Sigma}_{qR}^>(\omega) &= -i J_{qR}(\omega) [N_R(\omega) +1] e^{+i\hbar\omega\lambda_h} ,
\end{align}
with the Bose-Einstein distribution $N_R(\omega)=1/(e^{\beta_R\omega}-1)$. 
$\widetilde{F}_q$ is the electron-hole propagator containing spin current counting field $\lambda_s$, and the lesser and greater component in the energy domain from Eq.~\eqref{F_time} has the form \cite{Bijay}
\begin{align}
\widetilde{F}_q^< (\omega) =& -i2\pi g^2 \sum_{kk'} |\gamma_{k'\downarrow}|^2 
|\gamma_{k\uparrow}|^2 f_{L\downarrow}(\epsilon_{k'\downarrow})[1-f_{L\uparrow}(\epsilon_{k\uparrow})]  \notag \\
&\times e^{+i\lambda_s}\delta(\epsilon_{k'\downarrow}-\epsilon_{k\uparrow}-\omega)\delta(\omega-\omega_q) ,  \notag \\
\widetilde{F}_q^> (\omega) =& -i2\pi g^2 \sum_{kk'} |\gamma_{k'\downarrow}|^2 
|\gamma_{k\uparrow}|^2 f_{L\uparrow}(\epsilon_{k\uparrow}) [1-f_{L\downarrow}(\epsilon_{k'\downarrow})]  \notag \\
&\times e^{-i\lambda_s}\delta(\epsilon_{k'\downarrow}-\epsilon_{k\uparrow}-\omega) \delta(\omega-\omega_q) .
\end{align}

Since the electron-hole propagator $F$ is restricted by the delta functions $\delta(\epsilon_{k'\downarrow}-\epsilon_{k\uparrow}-\omega) \delta(\omega-\omega_q)$ due to energy conservation, one can drop the dependence of wave vector $q$ in $\underline{D}_q$, $\Sigma_{qR}$ and $F_q$ and let $\omega_q=\omega$, so that Eq. \eqref{ZZ1} can be rewritten as, 
\begin{equation}
{\cal G}(\lambda_s,\lambda_h) = -\int \frac{d\omega}{2\pi} \ln \det 
 \frac{\begin{bmatrix}
-\Sigma_{R}^{t}-F^{t}  &  \widetilde{\Sigma}_{R}^{<}+\widetilde{F}^{<} \\
\widetilde{\Sigma}_{R}^{>}+\widetilde{F}^{>} &  -\Sigma_{R}^{\bar{t}}-F^{\bar{t}} 
\end{bmatrix}  }
{\begin{bmatrix}
-\Sigma_{R}^{t}-F^{t}  &  \Sigma_{R}^{<}+ F^{<} \\
\Sigma_{R}^{>}+ F^{>} &  -\Sigma_{R}^{\bar{t}}-F^{\bar{t}} 
\end{bmatrix} } .
\end{equation}
Here the explicit dependence of the frequency $\omega$ on the Green’s functions, self-energies and electron-hole propagators is suppressed for notational simplicity.
Denoting the spin $\sigma$ electronic spectral densities in the left lead as $J_{L\sigma}(E)$, the electron-hole propagator could be written in integral forms
\begin{align}
\widetilde{F}^<(\omega) &= -i \int \frac{dE}{2\pi} \Big[ e^{+i\lambda_s} J_{\uparrow}(E_-) J_{\downarrow}(E_+) f_{\downarrow\rightarrow\uparrow}(E_+, E_-) \Big] , \notag \\
\widetilde{F}^>(\omega) &= -i \int \frac{dE}{2\pi} \Big[ e^{-i\lambda_s} J_{\uparrow}(E_-) J_{\downarrow}(E_+) f_{\uparrow\rightarrow\downarrow}(E_-, E_+) \Big] ,
\end{align}
with $E_{\pm}=E \pm \omega/2$, where we have used the short notation $ f_{\uparrow\rightarrow\downarrow}(E_-,E_+) = f_{L\uparrow}(E_-) [1- f_{L\downarrow}(E_+)] $ and $ f_{\downarrow\rightarrow\uparrow}(E_+,E_-) = f_{L\downarrow}(E_+) [1- f_{L\uparrow}(E_-)] $. 
A Lorentzian shaped form of electronic spectral density $J_{L\sigma}(E) = \frac{g\Gamma_\sigma(E)}{(E-\epsilon_\sigma)^2+\Gamma_\sigma(E)^2/4}$ is used in our numerical calculation. 
The summation of the time and anti-time ordered electron-hole propagator could be obtained from the identities, 
\begin{equation}
F^t(\omega) + F^{\bar t}(\omega) = F^<(\omega) + F^>(\omega) .
\end{equation}
From the relation to get retarded electron-hole propagator in the time domain,
\begin{equation}
F^r(t_1,t_2) =\theta(t_1-t_2)[F^>(t_1,t_2)-F^<(t_1,t_2)], 
\end{equation}
its expression in the energy domain is 
\begin{equation}
F^r(\omega)= i \int\frac{d\omega_1}{2\pi} \frac{F^>(\omega_1)-F^<(\omega_1)}{\omega -\omega_1 +i0^+},
\end{equation}
by using the identity $\int_0^\infty e^{iE t} dt=i/(E+i0^+)$.
$F^r(\omega)$ could be decomposed into $F^r(\omega) = {\rm Re} [F^r(\omega)]+i{\rm Im} [F^r(\omega)]$
with the imaginary and real parts
\begin{align}
{\rm Im} [F^r(\omega)] &= \frac{1}{2}[F^>(\omega)-F^<(\omega)] , \notag \\
{\rm Re} [F^r(\omega)] &=  P \int_{-\infty}^{+\infty} \frac{d\omega_1}{\pi} \frac{{\rm Im} [F^r(\omega)]}{\omega_1 -\omega} ,
\end{align}
by using the Plemelj formula $1/(E\pm i0^+)=P(1/E)\mp i\pi\delta(E)$. The difference between time and anti-time ordered electron-hole propagator could be obtained through \cite{Bijay}, 
\begin{equation}
F^t(\omega) - F^{\bar t}(\omega) = 2{\rm Re} [F^r(\omega)].
\end{equation}

	Using the Dyson equation of $D(\omega)$, Eq.~\eqref{Dyson}, one has
\begin{equation}
{\cal G}(\{\lambda\})= -\int \frac{d\omega}{2\pi} \ln \det \left[ 1+D
\begin{pmatrix}
0  &  \bar{\Sigma}_R^< + \bar{F}^< \\  \bar{\Sigma}_R^> +\bar{F}^> &  0
\end{pmatrix} \right] ,
\end{equation}
with $\{\lambda\}=\{\lambda_s, \lambda_h \}$, $\bar{\Sigma}_R =\widetilde{\Sigma}_R- \Sigma_R$ and $\bar{F} =\widetilde{F}- F$.
The Green's function $D(\omega)$ has the expression in the Keldysh space as,
\begin{equation}
D(\omega) = \frac{1}{[\Sigma_R^r + F^r][\Sigma_R^a + F^a]}
\begin{bmatrix}
\Sigma_R^{\bar{t}} + F^{\bar{t}}   &  \Sigma_R^< + F^< \\
\Sigma_R^> + F^> &  \Sigma_R^{t} + F^{t}
\end{bmatrix} ,
\end{equation}
by using the identities $F^{\bar{t}}+F^{t}=F^{<}+F^{>}$, $F^{t}=F^{<}+F^{r}$ and $F^{\bar{t}}=F^{<}-F^{a}$ and similar identities for $\Sigma_R$. The retarded magnonic self-energy is expressed as $\Sigma_R^r = -iJ_R(\omega)/2$. Without loss of generality, the magnonic reservoir spectral function can be expressed as \cite{Weiss}
\begin{equation}
J_R(\omega) = \pi\alpha\omega^s\omega_c^{1-s} e^{-\omega/\omega_c} ,
\end{equation}
where $\alpha$ is the dimensionless dissipation strength. The case $s>1$, $s=1$ and $s<1$ corresponds to super-Ohmic, Ohmic and sub-Ohmic dissipation, respectively. Different case of $s$ depends on the magnonic density of states. $\omega_c$ is the cut-off frequency of the magnonic reservoir.

	Further simplification enables us to have the expression of SCGF,
\begin{align} 
&{\cal G}(\lambda_s, \lambda_h) =  \notag \\
&-\int \frac{d\omega}{2\pi} \ln \left\{ 1 + \frac{\widetilde{\Sigma}_R^{<}\widetilde{F}^{>}+\widetilde{\Sigma}_R^{>}\widetilde{F}^{<}- \Sigma_R^< F^> - \Sigma_R^> F^< }{[\Sigma_R^r + F^r][\Sigma_R^a + F^a]} \right\},
\end{align}
which could also be expressed as
\begin{align} \label{ZZ2}
{\cal G}(\lambda_s, & \lambda_h) = -\int \frac{d\omega}{2\pi} \ln \bigg\{ 1-\int \frac{dE}{2\pi} A(E_-,E_+) \notag \\
\Big[ &(e^{+i\lambda_s+i\hbar\omega\lambda_h}-1)  f_{\downarrow\rightarrow\uparrow}(E_+, E_-) (N_R(\omega)+1)  \notag \\
 +&(e^{-i\lambda_s-i\hbar\omega\lambda_h}-1) f_{\uparrow\rightarrow\downarrow}(E_-, E_+) N_R(\omega) \Big] \bigg\},
\end{align}
with 
\begin{equation}
A(E_-,E_+) =\frac{J_R(\omega) J_{L\uparrow}(E_-)J_{L\downarrow}(E_+)}{|\Sigma_R^r(\omega)+F^r(\omega)|^2} .
\end{equation}
The first contribution of Eq. \eqref{ZZ2} describes that the spin-down electrons with energy $E_+$ flip to the spin-up states with energy $E_-$ via emitting a magnon with energy $\hbar\omega$. The second one describes a reverse process that the spin-up electrons with energy $E_-$ flip to the spin-down states with energy $E_+$ by absorbing a magnon carrying energy $\hbar\omega$.

	The spin current in the electronic reservoir and heat current in the magnonic reservoir could be derived by taking the first order derivative of ${\cal G}(\{\lambda\})$ with respect to $\lambda_s$ and $\lambda_h$, respectively at $\lambda_s=\lambda_h=0$, that is, $I_{s/h} =\partial {\cal G}(\{\lambda\}) /\partial (i\lambda_{s/h})|_{\lambda_s=\lambda_h=0}$. Then,
\begin{align}
&I_{s(h)} =  \int\frac{d\omega}{2\pi}\int\frac{dE}{2\pi}\Big\{ 
(\hbar\omega)^\alpha A(E_-,E_+) \times  \notag \\
& \left[ f_{\downarrow\rightarrow\uparrow}(E_+,E-) (N_R(\omega)+1) 
- f_{\uparrow\rightarrow\downarrow}(E_-,E_+) N_R(\omega)  \right] \Big\}  ,
\end{align}
where $\alpha=0$ for spin current $I_s$, and $\alpha=1$ for heat current $I_h$. The spin current noise can be obtained via $\langle\langle I_s^2\rangle\rangle =\partial^2 {\cal G}(\{\lambda\}) /\partial (i\lambda_s)^2|_{ \{\lambda\}=0 }$ and has the expression,
\begin{widetext}
\begin{align}
\langle\langle I_s^2\rangle\rangle = \int\frac{d\omega}{2\pi} \Bigg\{ &\bigg[\int\frac{dE}{2\pi} A(E_-,E_+) \big[ f_{\downarrow\rightarrow\uparrow}(E_+,E-) (N_R(\omega)+1) + f_{\uparrow\rightarrow\downarrow}(E_-,E_+) N_R(\omega) \big]\bigg]  \notag \\
+ &\bigg[\int\frac{dE}{2\pi} A(E_-,E_+) \big[ f_{\downarrow\rightarrow\uparrow}(E_+,E-) (N_R(\omega)+1) - f_{\uparrow\rightarrow\downarrow}(E_-,E_+) N_R(\omega) \big]\bigg]^2 \Bigg\} .
\end{align}
\end{widetext}

Since stochastic variables $\hat{I}_s$ and $\hat{I}_h$ fluctuate, the efficiency $\eta_s = \hat{I}_s\Delta\mu_s /\hat{I}_h$ defined as their ratios with $T_R>T_L$ fluctuate as well and can be characterized by the large deviation function (LDF). 
	LDF $J(\eta_s)$ is the exponential decay rate at which the probability to observe a given quantity $\eta_s$ during a long measurement time \cite{EF2, EF3},
\begin{equation}
P(\eta_s) =\lim_{t\rightarrow \infty} \exp\left[-J(\eta_s)t\right] . 
\end{equation}
It could be calculated through ${\cal G}(\lambda_s,\lambda_h)$ by setting $\lambda_s\rightarrow \lambda_s\Delta\mu_s$ and $\lambda_h=\eta_s\lambda_s$, and then minimizing ${\cal G}$ relative to $\lambda_s$, namely, \cite{Bijay, EF2, EF3}
\begin{equation}  \label{min_p}
J(\eta_s)=-\mathrm{min}_{\lambda_s}{\cal G}\left(\lambda_s\Delta\mu_s, \eta_s\lambda_s \right) .
\end{equation}